\documentclass[lettersize,journal]{IEEEtran}
\usepackage{amsmath,amsfonts}
\usepackage{algorithmic}
\usepackage{algorithm}
\usepackage{array}
\usepackage{textcomp}
\usepackage{stfloats}
\usepackage{url}
\usepackage{verbatim}
\usepackage{graphicx}
\usepackage{cite}
\usepackage{longtable}
\usepackage{enumerate}
\usepackage{atbegshi}%
\usepackage{xcolor}
\usepackage{hyperref}
\usepackage{booktabs}
\usepackage{multirow}
\usepackage{balance}
\usepackage{flushend}
\usepackage[flushleft]{threeparttable} 
\usepackage{tablefootnote}
\usepackage{soul} 
\usepackage{ulem} 
\usepackage{subcaption}
\usepackage{pifont}
\newcommand{\xmark}{\ding{55}}%

\usepackage{lipsum}

\usepackage{scalerel}
\usepackage{tikz}
\usetikzlibrary{svg.path}

\definecolor{orcidlogocol}{HTML}{A6CE39}
\tikzset{
  orcidlogo/.pic={
    \fill[orcidlogocol] svg{M256,128c0,70.7-57.3,128-128,128C57.3,256,0,198.7,0,128C0,57.3,57.3,0,128,0C198.7,0,256,57.3,256,128z};
    \fill[white] svg{M86.3,186.2H70.9V79.1h15.4v48.4V186.2z}
                 svg{M108.9,79.1h41.6c39.6,0,57,28.3,57,53.6c0,27.5-21.5,53.6-56.8,53.6h-41.8V79.1z M124.3,172.4h24.5c34.9,0,42.9-26.5,42.9-39.7c0-21.5-13.7-39.7-43.7-39.7h-23.7V172.4z}
                 svg{M88.7,56.8c0,5.5-4.5,10.1-10.1,10.1c-5.6,0-10.1-4.6-10.1-10.1c0-5.6,4.5-10.1,10.1-10.1C84.2,46.7,88.7,51.3,88.7,56.8z};
  }
}

\newcommand\orcidicon[1]{\href{https://orcid.org/#1}{\mbox{\scalerel*{
\begin{tikzpicture}[yscale=-1,transform shape]
\pic{orcidlogo};
\end{tikzpicture}
}{|}}}}

\begin{document}

\title{Synthetic Grid Generator: Synthesizing Large-Scale Power Distribution Grids using Open Street Map}

\author{Chandra Sekhar Charan Dande \orcidicon{0009-0008-7483-854X}, Luca Mattorolo \orcidicon{0009-0000-2532-5715}, Joel da Silva Andre \orcidicon{0000-0002-5956-9213}, Lydia Lavecchia \orcidicon{0009-0005-4574-2387}, Nikolaos Efkarpidis \orcidicon{0000-0001-8487-6824}\, and Damiano Toffanin \orcidicon{0009-0002-4255-3554}
\thanks{This paper was funded by Secure Switzerland AG.}
\thanks{(Corresponding author: Chandra Sekhar Charan Dande)}
\thanks{Chandra Sekhar Charan Dande, Luca Mattorolo, Joel da Silva André, Lydia Lavecchia, and Nikolaos Efkarpidis are with Secure Switzerland, 8048 Zürich, Switzerland.}
\thanks{Damiano Toffanin is with  University of St.Gallen (HSG), 9000 St. Gallen, Switzerland.}}



\maketitle

\begin{abstract}
Nowadays, various stakeholders involved in the analysis of electric power distribution grids face difficulties in the data acquisition related to the grid topology and parameters of grid assets. To mitigate the problem and possibly accelerate the accomplishment of grid studies without access to real data, we propose a novel approach for generating synthetic distribution grids (Syngrids) of (almost) arbitrary size replicating the characteristics of real medium- and low-voltage distribution networks. The method enables large-scale testing without incurring the burden of retrieving and pre-processing real-world data. The proposed algorithm exploits the publicly available information of Open Street Map (OSM). By leveraging geospatial data of real buildings and road networks, the approach allows to construct a Syngrid of chosen size with realistic topology and electrical parameters. It is shown that typical power-flow and short-circuit calculations can be performed on Syngrids ensuring convergence. Within the context of validating the effectiveness of the algorithm and the meaningful similarity of the output to real grids, the topological and electrical characteristics of a Syngrid are compared to their real-world counterparts. Finally, an open-source web platform named as Synthetic Grid Generator (SGG) and based on the proposed algorithm can be used by various stakeholders for the creation of synthetic grids. 
\end{abstract}


\section*{Nomenclature}
\subsection{Abbreviations}

\begin{tabular}{l l}
  COD & Coefficient of Overdimensioning\\
  CF   & Coincident Factor \\
  CRS & Coordinate Reference System\\
  DSO & Distribution System Operator\\
  EIA & Energy Information Administration \\
  HV & High Voltage\\
  LV & Low Voltage\\
  MV & Medium Voltage\\
  pu & per unit\\
  PV & Photovoltaics \\
  OSM & Open Street Map\\
  RES & Renewable Energy Sources\\
  Syngrid & Synthetic Distribution Grid\\
  SGG & Synthetic Grid Generator  
  
\end{tabular}

\subsection{Symbols}

\begin{tabular}{l l}
 $P_n$ & Active power of load $n$ in kW\\
 $P_\textrm{agg}$ & Aggregated power in kW\\
 $V_\textrm{op}$ & Cable operating voltage in kV\\
 $P_\textrm{cu}$ & Copper loss in kW\\
 $i_\textrm{m}$ & Maximum current in kA\\
 $P_\textrm{m}$ & Maximum power in kW\\
 $V_\textrm{n}$  & Nominal Voltage in kV\\
 $i_{1\phi}$ & Per phase current in kA\\
 $S_\textrm{r}$ & Rated apparent power in MVA\\
 $V_\textrm{k}$ & Short-circuit voltage in kV\\
  
\end{tabular}

\section{Introduction}

\IEEEPARstart{I}{n} recent years, power systems have been subjected to various dynamics due to the increasing penetration of distributed energy resources (DERs), microgrids, and responsive loads~\cite{HPadullaparti}. This situation results in unstable electricity prices, consequently, diminishing further investments~\cite{TSELIKA}. To tackle these challenges, it is essential to expand and develop the electricity grid and network infrastructure. This setup allows for efficient electricity transfer from regions with excess power to those experiencing deficits, thereby ensuring supply stability. With European growing dependence on fluctuating renewable energy, the infrastructure significance is heightened, ensuring it remains a critical area of research likely to attract increasing attention in the future~\cite{QChunzi}. The availability of real distribution grids is crucial for grid analyses and grid planning studies. Due to security and privacy reasons, distribution grid models with complete information are not publicly accessible. Furthermore, there is lack of information due to unnotified changes, network reconfigurations after faults or maintenance operations, and outdated grids that have never been documented~\cite{LuanWen,PappuSat}. As a result of them, there is an increasing interest in developing synthetic grids that can mimic real grid conditions, enabling various grid-related studies and future forecasts.

There are multiple applications for synthetic grids, some of which are described below. For the purposes of researching and analyzing distribution grids, relevant grid models are essential for research groups and academic institutes. Another application pertains to energy suppliers, who aim to assess their business models but lack access to grid data, as this information is owned by distribution system operators (DSO)~\cite{nefkarpidis}. A third use case involves DSOs that may not have comprehensive data on outdated distribution grids; thus, synthetic grids could accurately represent real grids with a sufficient accuracy. It is evident from the aforesaid instances that the availability of grid data is required for the accomplishment of business processes and technical studies at different domains of interest.  

The extant literature explores the challenge of the limited availability of distribution network data for grid studies. For instance, the work presented in~\cite{Trapovski8542054} builds a synthetic distribution grid for load allocation and planning by leveraging publicly accessible data. This methodology focuses on optimization of power distribution system planning but is tailored to individual cases, and necessitates the collection of specific data for each area to fabricate a synthetic grid. Moreover, a framework for generating a synthetic grid suitable for dynamic analyses is proposed in~\cite{AButtner10.1063/5.0155971}. However, their approach encounters issues with power flow convergence when the number of grid nodes surpasses 10,000. The absence of geographical information in the transmission line drawings further results in inaccuracies in line length. In~\cite{SJYoung8585792}, the authors disassemble existing proprietary power systems and recompose them from disparate components to create a realistic synthetic power system. A synthetic network for the United States is devised in~\cite{KMGegner7459256}, with load placements derived from census data and generator selections based on Energy Information Administration (EIA) data. Nevertheless, critical analyses such as power flow convergence are not addressed. Table~\ref{tab:litGaps} elucidates the gaps identified in the literature, revealing that the existing studies do not completely cover the features required for the generation of synthetic grids. It can also be observed that the majority of the current literature does not incorporate geographical and spatial assessments. Moreover, most methodologies do not use time-series data for the generation of synthetic profiles, therefore, they are inappropriate for the conduction of time-series simulations. Additionally, certain methodologies lack electrical analysis, which is crucial for the operation of electrical grids. Various studies also utilize actual grid data, which limits the algorithm to specific regions. From the literature review, it is also clear that the existing approaches can rarely be extended to any size limiting the generation of large-scale grids. Furthermore, the development of a web application is crucial for industrial stakeholders, as the practical use of the algorithm is often restricted by the time required to replicate it. A web application would enhance usability and broaden the applicability of the algorithm.

\begin{table*}[t!]
    \centering
     \caption{Gaps in State-of-the-Art Methodologies}
    \begin{tabular}{|c|c|c|c|c|c|c|c|c|c|}
    \hline
     Feature  & \cite{Trapovski8542054} & \cite{AButtner10.1063/5.0155971} & \cite{SJYoung8585792} & \cite{KMGegner7459256} & \cite{ZWang} &  \cite{Elyas} & \cite{Eschweitzer} & \cite{Jhu} & \cite{ABirchfield}\\
    \hline
      Geographical and spatial evaluation   & \checkmark & \xmark & \xmark & \checkmark & \xmark & \xmark & \checkmark & \xmark & \xmark \\
      \hline
      Generation of synthetic profiles using time-series data & \xmark & \checkmark & \xmark & \xmark & \xmark & \xmark & \xmark & \xmark & \xmark\\
      \hline
      Conduction of load flow analysis and stress tests & \xmark & \checkmark & \checkmark & \xmark & \xmark & \xmark & \xmark & \checkmark & \checkmark \\
      \hline
      No use of real-world data (not case specific) & \xmark & \checkmark & \xmark & \checkmark & \xmark & \xmark & \xmark & \xmark & \xmark\\
      \hline
      Scalable to any size & \xmark & \xmark & \xmark & \xmark & \checkmark & \checkmark & \xmark & \xmark & \xmark \\
      \hline
      Available web application & \xmark & \xmark & \xmark & \xmark & \xmark & \xmark & \xmark & \xmark & \xmark \\
      \hline
    \end{tabular}
    \label{tab:litGaps}
\end{table*}

\par In this paper, we introduce a novel method for generating synthetic grids, herein referred to as Syngrid, whenever synthesized by the proposed algorithm. The Syngrid is meticulously constructed utilizing Open Street Map (OSM) data, encompassing the roadways and building infrastructures. A Syngrid can be instantiated for any geographic region with an appropriate coordinate reference system (CRS) and is scalable to any desired number of nodes. This process is accomplished without reliance on any publicly available grid data. Furthermore, during algorithm development, comprehensive geographical and spatial considerations ensure that the cable lengths are commensurate with real-world grid dimensions. Notably, electrical convergence is rigorously maintained throughout the examined period in case of time-series simulations, eschewing the use of actual grid data. Additionally, various stakeholders from different domains, e.g., industry, research and academy, can use the Synthetic Grid Generator (SGG) application without requiring any expertise on the design of distribution grids~\cite{Syngrid}. Particularly, the application facilitates the generation of new synthetic grids or utilization of pregenerated instances without necessitating the algorithm reconstruction from the first principles. To the authors' best knowledge, this work represents the inaugural open-source application dedicated to the generation of synthetic distribution grids, with the underlying algorithm effectively addressing the deficiencies prevalent in the extant literature. As a whole, the novelty of this study can be summarized as follows:

\begin{enumerate}
    \item The algorithm can be extended to any number of nodes enabling the generation of large-scale distribution grids. 
    \item Time-series profiles representing the end-customers' energy demand are also available with the grid.
    \item There is no utilization of data from real power grids.
    \item The grid that is generated is always electrically appropriate. Grid analysis, such as power flow and short-circuit calculations, can be performed. 
    \item An web platform is available to various stakeholders for creating new grids or downloading existing ones without the need to build the algorithm from the ground.
\end{enumerate}

The remainder of the paper is organized into four sections. Section 2 explains the algorithm used for generating Syngrids. Section 3 presents the results highlighting the comparative analysis of static characteristics between a real grid and a Syngrid. Furthermore, electrical analysis on the generated Syngrid is also provided. Finally, the last section draws conclusions based on our findings.


\begin{figure}[b!]
\centering
\includegraphics[width=0.5\textwidth]{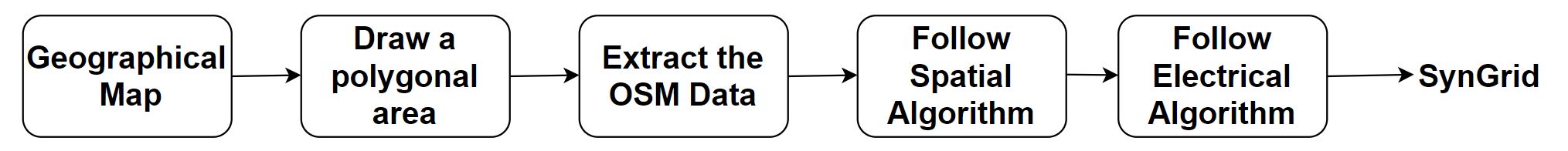}
\caption{Flowchart of the algorithm.}
\label{fig:flowchart}
\end{figure}

\section{SGG Methodology}

Figure~\ref{fig:flowchart} illustrates the Syngrid generation process. The process begins with marking a polygonal area on a geographical map. Next, OSM data is extracted for the specified polygonal area. Subsequently, the spatial algorithm is employed to process the geographical data. This is then complemented by the electrical algorithm to generate the Syngrid as the final output.

\begin{figure}[t!]
\centering
\includegraphics[width=0.45\textwidth]{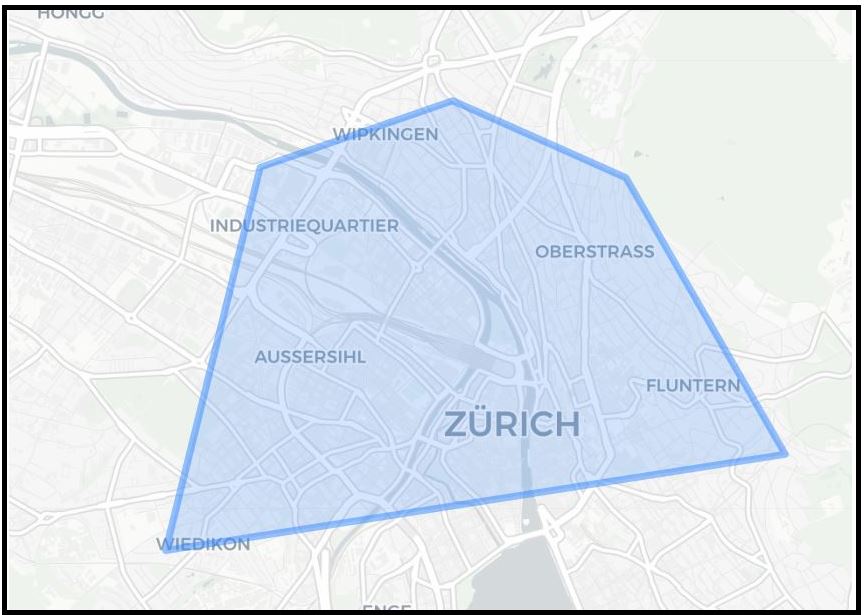}
\caption{A geographical map where a polygon is drawn to make the Syngrid.}
\label{fig:geom}
\end{figure}

\begin{figure}[b!]
\centering
\includegraphics[width=0.5\textwidth]{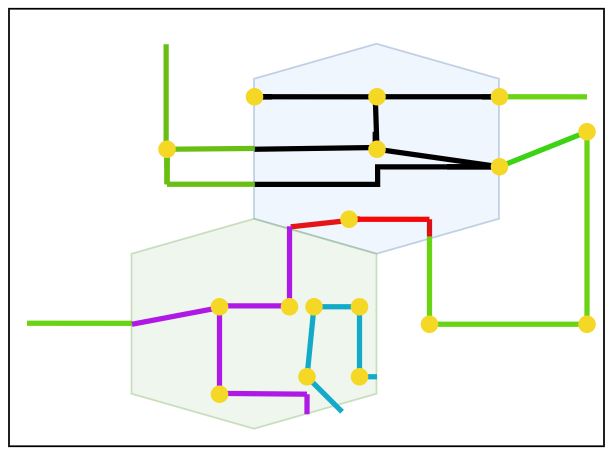}
\caption{Visualization of two hexagonal polytopes, where a LV grid is spanned in each of them.}
\label{fig:tesselation}
\end{figure}

\subsection{Spatial Algorithm}
\begin{figure}[t!]
\centering
\includegraphics[width=0.48\textwidth]{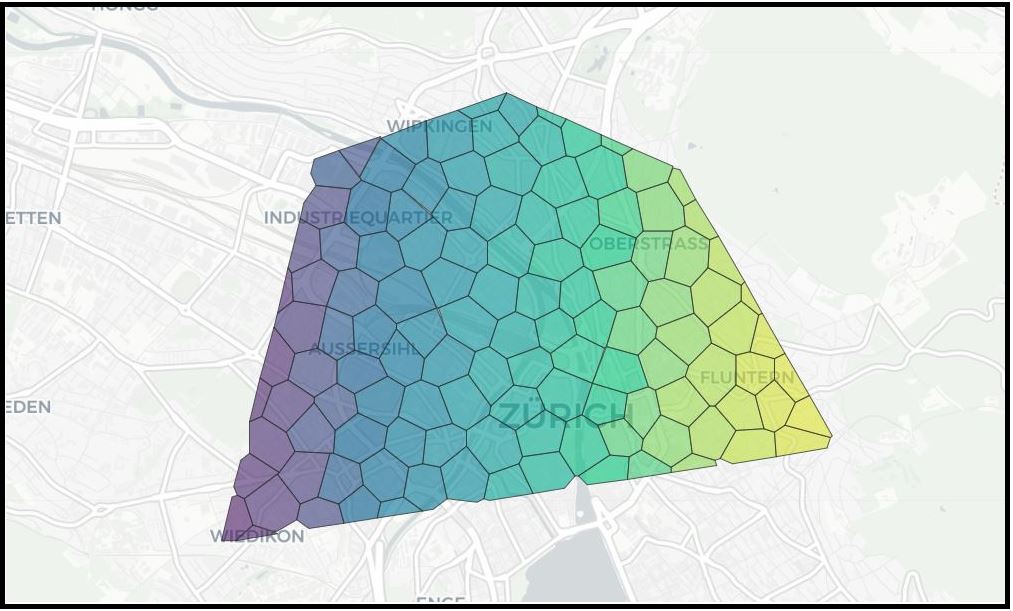}
\caption{Depiction of predefined polygon transformed into hexagonal polytopes to facilitate the formation of the LV grid within each polytope.}
\label{fig:tesselationonpolygon}
\end{figure}

In the initial phase of building a Syngrid, a polygonal area is chosen, as depicted in Figure~\ref{fig:geom} on a geographical map. Firstly, the identified polygonal area serves as the focal point for the grid's development. Next, OSM data pertinent to the selected region must be retrieved to provide essential geographical information. Subsequently, the polygonal area is subdivided into polytopes, each characterized by a predetermined average radius, as illustrated in Figure~\ref{fig:tesselation}. The polytopes serve as the basis for delineating the grid boundaries and accommodating individual Low Voltage (LV) grids within each partition. To streamline the process, the map is cropped in accordance with the polytopes, facilitating focused consideration of each tessellation for the creation of the LV grid infrastructure as illustrated in Figure~\ref{fig:tesselationonpolygon}. This meticulous approach ensures systematic planning and execution of the Syngrid within the specified geographical area.

\subsection{Generation of LV Grids}
In the process of establishing a LV grid within each polytope region, several crucial steps are taken. Firstly, connections are established from buildings to the nearest roads within each polytope, serving as the interface between the LV network and the loads. Residential load allocation within buildings is then determined based on their cross-sectional area, with an assumption made regarding the number of consumers per square meter, each assigned a specific load value. Furthermore, to integrate photovoltaic (PV) systems distributed throughout the grid, a penetration rate of 10\% is assumed, with 10\% of the loads (selected randomly) also including PV generation at 50\% of their respective power demand. In the future, we will also incorporate the capability for users to modify the PV penetration level.

Moreover, the network is transformed into a spatial network by splitting intersectional roads into separate edges, ensuring that each road, which represents an edge in the graph, is delineated with its associated geometry. Focus is directed towards the most interconnected component in the resulting graph, disregarding others to ensure efficient load supply through a single feeder in each LV grid. Given the radial nature of LV grids, this property is enforced by pruning the graph and connecting loads to the root node via the shortest path.

The root node, designated as the connection point for the distribution transformer, serves as a pivotal component in the grid architecture. A transformer is also added as an edge to the principal node to facilitate connection to the Medium Voltage (MV) grid. Electrical parameters are then assigned to the resulting LV grid utilizing an electrical algorithm explained in subsection~\ref{Electrical Algorithm}. This comprehensive procedure is iterated for each polytope, ensuring the systematic generation of a LV grid across the designated geographical area.

\begin{figure}[t!]
\centering
\includegraphics[width=0.5\textwidth]{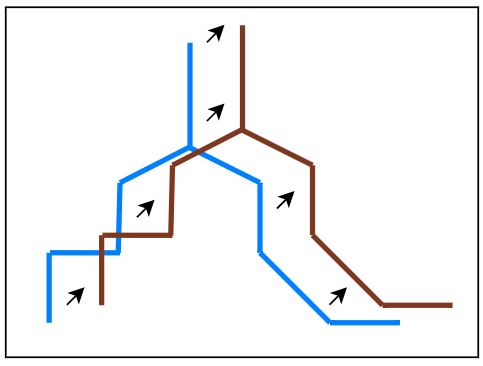}
\caption{Translation of road network to prevent any overlapping and improve grid visualization.}
\label{fig:translation}
\end{figure}

\subsection{Generation of MV Grid}
To initiate the generation of the MV grid, a re-evaluation of the raw OSM data is necessary. Once all LV grids have been established, the MV grid can be developed. To ensure clear visualization and prevent overlap between the MV and LV grids, the MV grid is generated based on a translated road network, as highlighted in Figure~\ref{fig:translation}. Geometric connections are then established between the MV nodes originating from the MV/LV transformers and the nearest roads in the translated network. This connectivity is represented graphically and subjected to a similar procedure as applied in the creation of the LV grid. Particularly, the most connected node, or root node, is identified, and the graph is pruned to ensure that the MV nodes are linked to this root node. 

 Subsequently, electrical parameters are assigned to the resulting MV grid utilizing an electrical algorithm explained in subsection~\ref{Electrical Algorithm}. To integrate the MV grid into the broader electrical infrastructure and facilitate its interconnection to the high voltage (HV) network, a HV/MV power transformer is connected to the root node in the MV grid, serving as a feeder to the MV grid. This systematic approach ensures the efficient and effective generation and integration of the MV grid within the designated geographical area, contributing to the overall functionality and reliability of the synthetic electrical grid system.

\subsection{Electrical Algorithm}
\label{Electrical Algorithm}
Upon the completion of grid generation using the spatial algorithm, the subsequent task involves the assignment of electrical parameters to ensure its functionality. Initially, the algorithm is applied to a single LV grid, with plans for iterative application to the remaining LV grids and the MV grid. This approach is adopted because of the algorithm compatibility with single transformer configurations, which align well with distribution grids. Within the selected LV grid, load aggregation is performed from the child nodes to the parent node, in order to calculate the total load in the radial grid. Concurrently, the summarized weights of the end nodes are mapped to their corresponding lines (edges).

For each line, dimensioning of the power cable requires an estimation of the maximum expected load. The maximum load can then be estimated from the number of customers served by the line. All the consumers are assumed to be homogeneous with the same nominal power, and they have different consumption behaviors which were taken from~\cite{lpg}. In this paper, we utilize a function that estimates the maximum expected apparent power with respect to the number of consumers.

\begin{figure}[t!]
    \centering
    \includegraphics[width=0.5\textwidth]{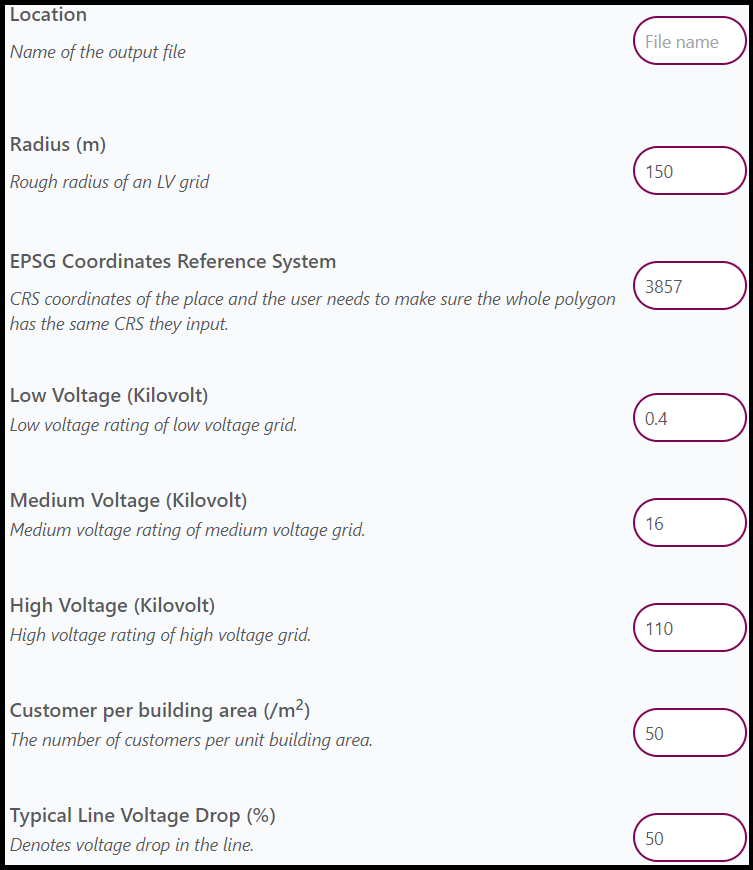}
    \caption{Synthetic Grid Generator customizable parameters}
    \label{fig:SGGPara}
\end{figure}

For a set of $n$ homogeneous consumers with rated apparent power $S_\textrm{r}$, the maximum aggregated load $P_\textrm{m}$ is defined as 

\begin{equation}
    P_\textrm{m} = n \cdot S_{r} \cdot \textrm{CF}(n),
\end{equation}
where $\textrm{CF}(n)$ is the coincidence factor function, which is defined as the ratio of the peak of the aggregated load to the sum of peaks of individual loads according to~\cite{damiano}, as follows:

\begin{equation}
    \textrm{CF}(n) = \frac{\max(P_\textrm{agg})}{\Sigma \max (P_n)}
\end{equation}

We build an estimation of $\textrm{CF}(n)$ via Monte Carlo sampling of a pool of load time series. Further explanation of the coincident factor can be seen in the Appendix.



 Given the in prior assumed rated and base voltage $V_n$ for all LV and MV grids, the per-phase current of each line can be computed using 
\begin{equation}
    i_{1\phi} = \frac{P_\textrm{m}}{\sqrt{3} \cdot V_\textrm{n}}.
\end{equation}

The practical cable parameters are then assigned to each line based on the following conditions:
\begin{equation}
    0.5 \cdot V_\textrm{n} \leq V_\textrm{op} \leq 1.5 \cdot V_\textrm{n},
\end{equation}
    \begin{equation}
        i_\textrm{m} \geq \textrm{COD} \cdot i_{1\phi},
    \end{equation}
ensuring appropriate electrical characteristics.

Subsequently, all lines are defined with the requisite electrical parameters. The subsequent step involves assigning parameters to the transformer within the grid (MV/LV transformer for LV grid and HV/MV transformer for MV grid). Parameters, including the rated apparent power $S_\textrm{r}$, short-circuit voltage $V_\textrm{k}$, and copper losses $P_\textrm{Cu}$, are ascertained based on the maximum power. These parameters are computed utilizing empirical formulas derived from \cite{abblibrary_2024}\cite{siemmens}, as follows::
 \begin{equation}
     S_\textrm{r} = P_\textrm{m} \cdot \textrm{COD},
 \end{equation}
    \begin{equation}
        P_\textrm{Cu} = 1.5+\frac{-0.7}{3}\log_{10}(10 \cdot S_\textrm{r}),
    \end{equation}
    \begin{equation}
        V_\textrm{k} = 4+\frac{8}{3}\log_{10}(10 \cdot S_\textrm{r}).
    \end{equation}
where $P_\textrm{m}$ is the maximum power in kW, COD is the coefficient of overdimensioning and $S_\textrm{MVA}$ is the apparent power in MVA. Upon computing these values, electrical parameters are assigned to all lines and transformers within the considered grid, ensuring comprehensive functionality and compatibility with the broader electrical infrastructure.

Figure \ref{fig:SGGPara} delineates the suite of customizable parameters within the open-source web platform Synthetic Grid Generator (SGG). Initially, the parameter
`location' is presented, allowing for arbitrary user-defined settings. Subsequently, the `radius' parameter governs the spatial extent of each LV grid. Specifically, the overarching polygon is subdivided into multiple polytopes for LV grid formation by this parameter. Following this, the EPSG Coordinate Reference System (CRS) is introduced as the most critical attribute; an incorrect CRS can hinder the Syngrid generation algorithm. It is imperative that the selected polygon is entirely encompassed within a single CRS, as this parameter facilitates the extraction of road and building data from OSM. The parameters for LV, MV, and HV are then outlined, requiring user input corresponding to the respective voltage levels of the syngrid. Additionally, the 'customers per building area' parameter determines the allocation rate of consumers per specified area. The standard default value is 50, implying that one consumer is assigned for every 50 square meters. For instance, a structure spanning 300 square meters would correspondingly have 6 consumers assigned. Finally, the `Typical Line Voltage Drop' parameter is required, signifying the voltage drop across a line per length, which is instrumental for determining transformer profiles.

\section{Results}
\subsection{Statistical Characteristics}
In order to compare the characteristics of the real grid (Realgrid) and the Syngrid, we selected the following parameters to represent the similarity between their topologies:
\begin{itemize}
    \item Number of customers per km: This ratio is calculated as the total number of customers divided by the total length spanned by the grid.
    \item Number of LV grids: This parameter indicates the number of MV/LV transformers in the grid.
    \item Average diameter of all LV grids: This is the average across all LV grids of the maximum distances between any two vertices within each LV grid.
\end{itemize}

\begin{table}[t!]
    \centering
    \caption{Comparison of Statistical Characteristics Between Realgrid and Syngrid}
    \begin{tabular}{c|c|c}
    \midrule
    Quantity & Realgrid & Syngrid \\
    \midrule
    Number of customers per km & 29 & 31 \\
    Number of LV grids & 152 & 167 \\
    Average diameter of all the LV grids & 0.653 km & 0.538 km \\
    \bottomrule
    \end{tabular}
    \label{tab:example}
\end{table}

\begin{table}[b!]
\caption{\textrm{Summary of Power Flow Results}}
\centering
\begin{tabular}{l|l|l|l}
\toprule
\textrm{Type} & \textrm{Voltage level} & \textrm{Active power} & \textrm{Reactive power} \\
\midrule
\multirow{3}{*}{\textrm{Load}} & \textrm{16.0 kV} & \textrm{0 MW} & \textrm{0.00 var} \\ 
& \textrm{110.0 kV} & \textrm{0.00 MW} & \textrm{0.00 var} \\ 
& \textrm{0.4 kV} & \textrm{16.23 MW} & \textrm{0.00 var} \\ 
& \textrm{All levels} & \textrm{16.23 MW} & \textrm{0.00 var} \\ 
\midrule
\multirow{3}{*}{\textrm{Generation}} & \textrm{16.0 kV} & \textrm{0.00 kW} & \textrm{0.00 var} \\ 	
& \textrm{110.0 kV} & \textrm{0.00 kW} & \textrm{0.00 var} \\ 
& \textrm{0.4 kV} & \textrm{879.80 kW} & \textrm{0.00 var} \\ 
& \textrm{All levels} & \textrm{879.80 kW} & \textrm{0.00 var} \\
\midrule
\multirow{3}{*}{\textrm{Flow}} & \textrm{16.0/0.4 kV} & \textrm{15.65 MW} & \textrm{117.19 kvar} \\ 	
& \textrm{16.0 kV} & \textrm{827.65 MW} & \textrm{-28.25 Mvar} \\ 
& \textrm{0.4 kV} & \textrm{74.47 MW} & \textrm{225.87 kvar} \\ 
& \textrm{110.0/16.0 kV} & \textrm{15.78 MW} & \textrm{-667.23 kvar} \\
& \textrm{All levels} & \textrm{933.54 MW} & \textrm{-28.58 Mvar} \\
\midrule
\multirow{3}{*}{\textrm{Losses}} & \textrm{16.0/0.4 kV} & \textrm{85.69 kW} & \textrm{323.66 kvar} \\ 	
& \textrm{16.0 kV} & \textrm{43.07 kW} & \textrm{-1.11 Mvar} \\ 
& \textrm{0.4 kV} & \textrm{299.03 kW} & \textrm{117.19 kvar} \\ 
& \textrm{110.0/16.0 kV} & \textrm{55.00 kW} & \textrm{670.03 kvar} \\
& \textrm{All levels} & \textrm{482.80 kW} & \textrm{2.80 kvar} \\
\bottomrule
\end{tabular}
\label{table:powerflow_results}
\end{table} 

\subsection{Statistical Analysis}
A Realgrid obtained from a distribution system operator (DSO) in Switzerland is considered for comparison. This grid spans the central part of Switzerland near the city of Bern and supplies electricity to approximately 11,000 customers.

A Syngrid has been generated for the same geographical area as the Realgrid. To ensure a fair comparison, Syngrid is created with an equivalent number of customers as Realgrid. 

As explained in the reference \cite{ABEYSINGH}, topological parameters are a key to characterize different electricity distribution networks. It can be noticed in the comparison analysis in Table \ref{tab:example} that the topological characteristics of Syngrid, generated in a geographical area similar to Realgrid, are comparable with the topological features of Realgrid. 



\begin{figure}[t!]
\centering
\includegraphics[width=0.5\textwidth]{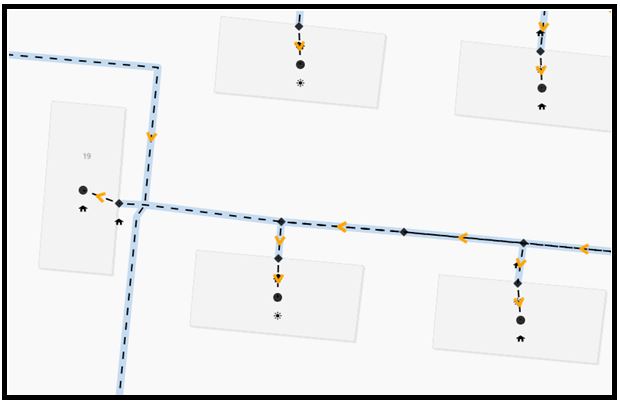}
\caption{Pictorial representation of active power flow to various loads.}
\label{fig:powerflowpictorial}
\end{figure}

\subsection{Electrical Analysis}
Syngrid, previously generated for the geographical location of Realgrid, is now under consideration for conducting electrical analyses, such as power flow and short-circuit power analysis. To provide an overview of the electrical characteristics of the grid, it comprises 986 buses and 168 transformers.

All electrical analyses were conducted using the software Adaptricity, which also facilitates protection analysis, stress-test simulation, hosting capacity calculation, N-1 analysis, time-series simulations, and power quality assessment \cite{Adaptricity}. All aforesaid analyses can be performed on Syngrid, ensuring convergence.




\begin{table}[b!]
    \centering
    \caption{Results From a Short Circuit Analysis}
    \begin{tabular}{l|m{1.2cm}|m{1.5cm}|m{1.8cm}}
    \toprule
      BusID   & Subgrid grounding & Minimum current [kA] & Maximum peak current [kA] \\
    \midrule
    Bus 1 & Direct& 2.497 & 8.100\\ 
    Bus 2 & Direct & 0.833 & 2.161 \\
    Bus 3 & Direct & 0.824 & 2.138 \\
    Bus 4 & Direct & 4.618 & 30.499\\
    Bus 5 & Direct & 8.183 & 30.363\\
    Bus 6 & Direct & 2.932 & 10.438\\
    Bus 7 & Direct & 4.138 & 21.698\\
    Bus 8 & Direct & 1.789 & 5.150\\
    Bus 9 & Direct & 2.325 & 7.253\\
    Bus 10 & Direct & 1.206 & 3.241\\
     \bottomrule   
    \end{tabular}
    
    \label{tab:shortcktanalysis}
\end{table}

\begin{figure}[t!]
\centering
\subfloat[][]{%
\label{fig:boxplotsa}%
\includegraphics[width=\columnwidth]{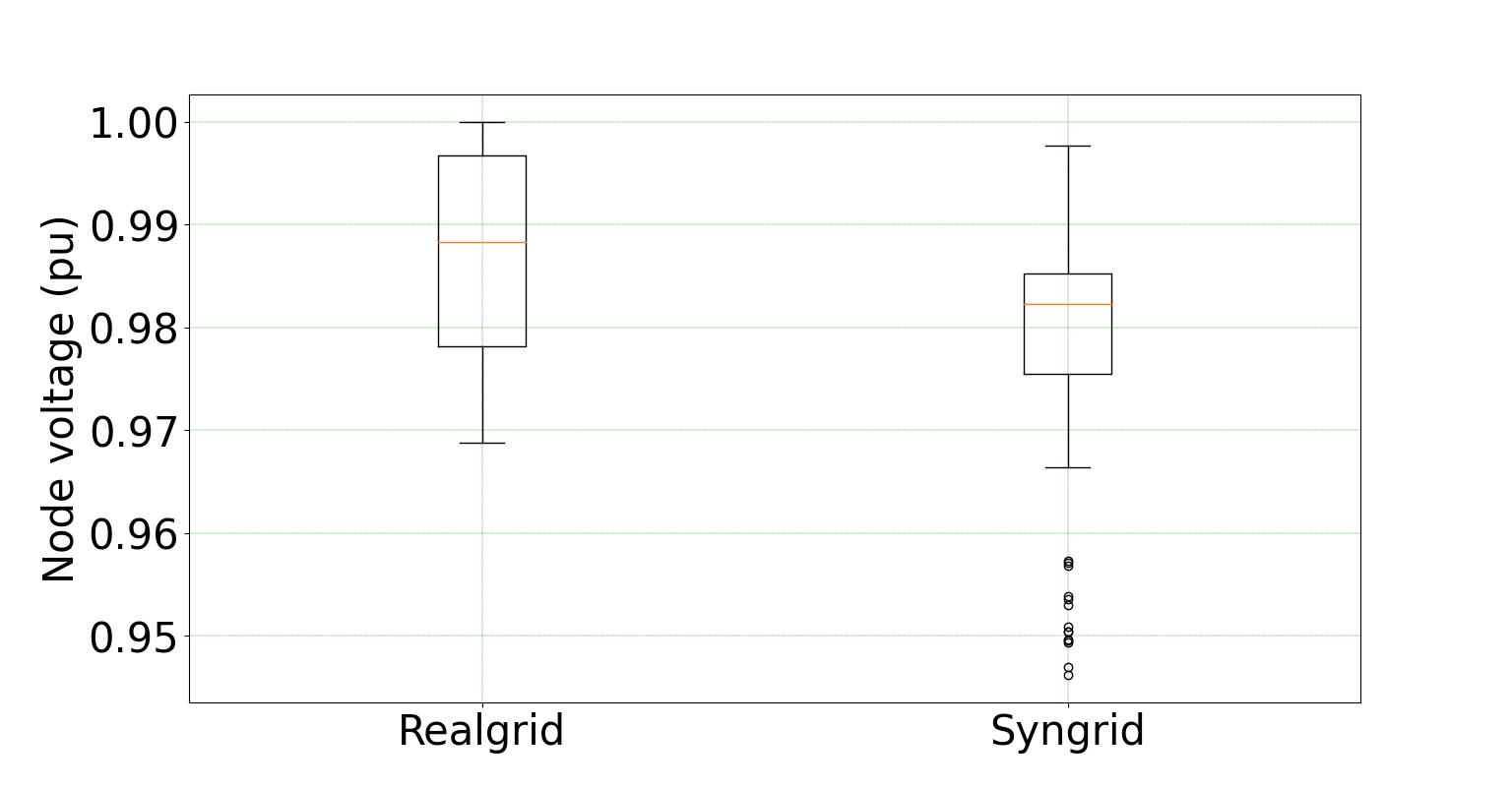}}
\hfill
\subfloat[][]{%
\label{fig:boxplotsb}%
\includegraphics[width=\columnwidth]{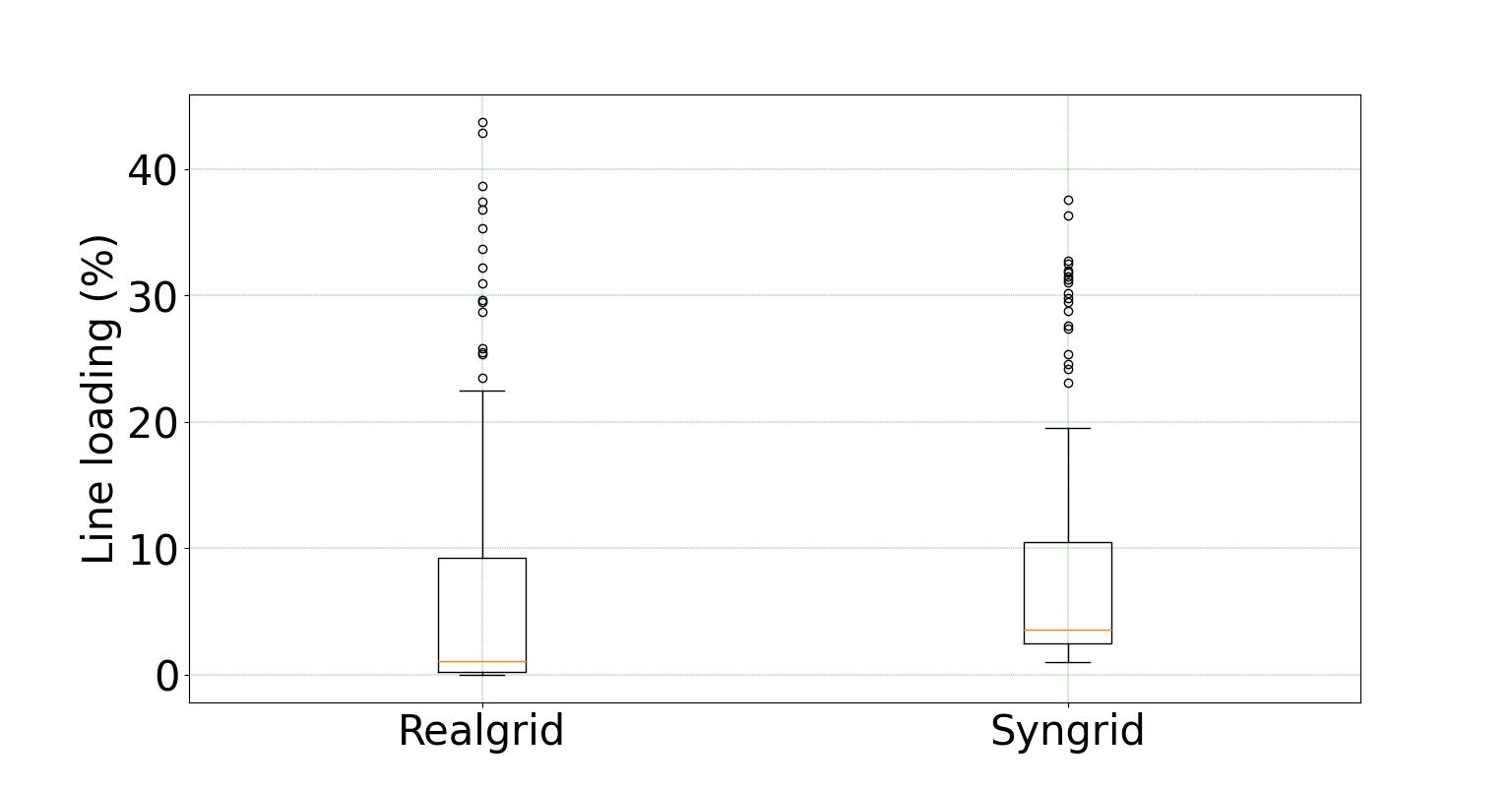}}
\caption[Optpeak]{\textrm{Comparison of electrical analysis between Realgrid and the corresponding Syngrid based on box plots of:} 
\subref{fig:boxplotsa} \textrm{voltage magnitude of all buses, and}  
\subref{fig:boxplotsb} \textrm{and loading of lines.} 
}%
\label{fig:boxplots}%
\end{figure}

Table~\ref{table:powerflow_results} presents a detailed exposition of the power flow analysis results. It is evident that the loads and generation are confined to the lower distribution level of 0.4 kV, characteristic of the current iteration of the proposed algorithm. Should the need arise, the algorithm can also be extended to accommodate loads at different voltage levels. It is further observed that reactive power flow and reactive power losses are prevalent across all distribution levels of the grid, enhancing the simulation fidelity and more accurately reflecting the characteristics of a real-world grid. 

Figure~\ref{fig:powerflowpictorial} provides a schematic depiction of power flow through lines to discrete loads. Additionally, it illustrates the methodology in establishing connections with each building, demonstrating the convergence of power flow, with arrows denoting the direction of power flow. In this manner, the connection points of buildings and the distribution lines can be exploited through the dynamic visualization of power flow in the grid. 


Table~\ref{tab:shortcktanalysis} delineates selected bus identification numbers along with their corresponding minimum and maximum peak current levels during short-circuit scenarios. In the context of the entire grid, the peak current maxima spans from 0.418 to 247.66~kA, while the minima ranges from 0.169 to 27.809~kA.

Next, a comparative analysis is conducted between the Realgrid and the Syngrid created considering both the voltage magnitudes of all buses and the thermal loading of all lines. For the comparison, typical box plots are used displaying the minimum, median, maximum values, 25\% and 75\% quartiles, as well as any outliers. 
Figure~\ref{fig:boxplotsa} compares the box plots of node voltage magnitudes in per unit (pu) values for several transformers between the Realgrid and their counterparts in Syngrid. It is evident that the median values of node voltages are approximately equivalent, while the minimum and maximum voltages considerably coincide excluding the outliers of Syngrid. Figure~\ref{fig:boxplotsb} includes a comparison of the relative thermal loading for the distribution lines of the aforementioned grids. Similarly to the node voltages, the line loadings mainly range from approximately 0\% to 10\% with maximum values of around 20\% for both grids. Moreover, the outliers indicate considerably similar distribution from about 23\% to around 40\%. It can be concluded from this analysis that the Syngrid can reproduce with high accuracy the status of real grids, which can be used for further grid studies without the use of actual grid data.

\section{Conclusion}
We propose a novel approach for generating Syngrid from OSM data. Additionally, we introduce an open-source platform, SGG, enabling various stakeholders to generate synthetic grids that share similarities with the real grid. Comparative analysis of the statistical characteristics of a Swiss grid and Syngrid, both situated in the same geographical location, confirms notable similarities. Moreover, we conduct electrical analyses, including power flow and short circuit analysis for illustrative purposes and evaluation of results; other analyses can also be performed. From the assessment, it is evident that the Syngrid can derive similar results of node voltages and line loadings with real grids. Finally, the flexibility of Syngrid allows it to scale to any desired number of buses, ensuring reliable convergence for all electrical analyses.

\section*{Acknowledgments}
We thank Secure Switzerland AG for funding and providing the resources for the accomplishment of this study. We also thank Energie Seeland AG, DSO of Lyss area, for the provision of grid information.

\appendix[Estimation of Coincidence factor function]
To build an approximation of $\textrm{CF}(n)$, we estimate its value on log-spaced values of $n$ and then interpolate the results. We considered values of $n_i = 2^i$ with $i=\{1,2,..,6\}$. After $n_6=64$, we assume an asymptotically flat behavior and set $\textrm{CF}(n)=\textrm{CF}(6)$ $\forall$ $n \geq 64$.

To compute the value of $\textrm{CF}(n_i)$ we sample $n_i$ time-series, compute the aggregated time-series, and select the maximum value. Repetition of the process $k_i$ times is required, we obtain a set of aggregated peak powers $k_i$ $\textrm{P}_j^{max}$, with $j=\{1,..,n\}$. We then define each $\textrm{CF}(n_i)$ as:

\begin{equation}
    \textrm{CF}(n_i) = \frac{1}{S_\textrm{r}}\frac{\max \{P_j ^{max}\}}{n_i}.
\end{equation}



%

\bibliographystyle{IEEEtran}
\bibliography{main} 


\section{Biography Section}

\vskip -2\baselineskip plus -1fil

\begin{IEEEbiography}
[{\includegraphics[width=1.02in,height=1.2in,clip]{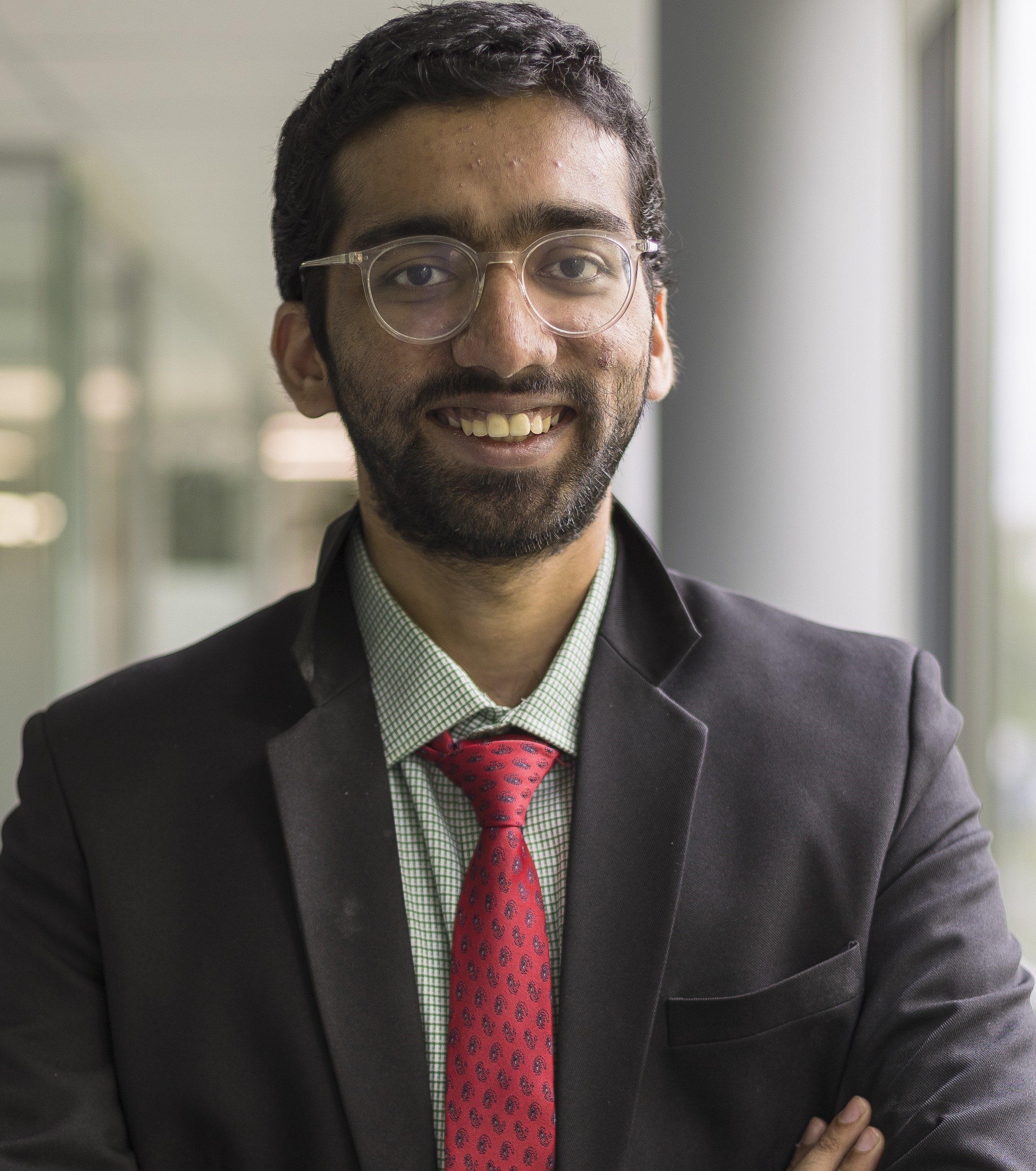}}]{Chandra Sekhar Charan Dande} fulfilled his Master's studies in Energy Science and Technology under the Department of Information Technology and Electrical Engineering at ETH Zürich, in 2024. He is currently working as a part-time Project Engineer at Secure Switzerland AG. Soon, he will be starting his PhD in the Institute of Automation of Complex Power Systems at RWTH Aachen. His main interests lie in the dynamics and control of the power system. 
\end{IEEEbiography}

\vskip -2\baselineskip plus -1fil

\begin{IEEEbiography}
[{\includegraphics[width=1.03in,height=1.25in,clip]{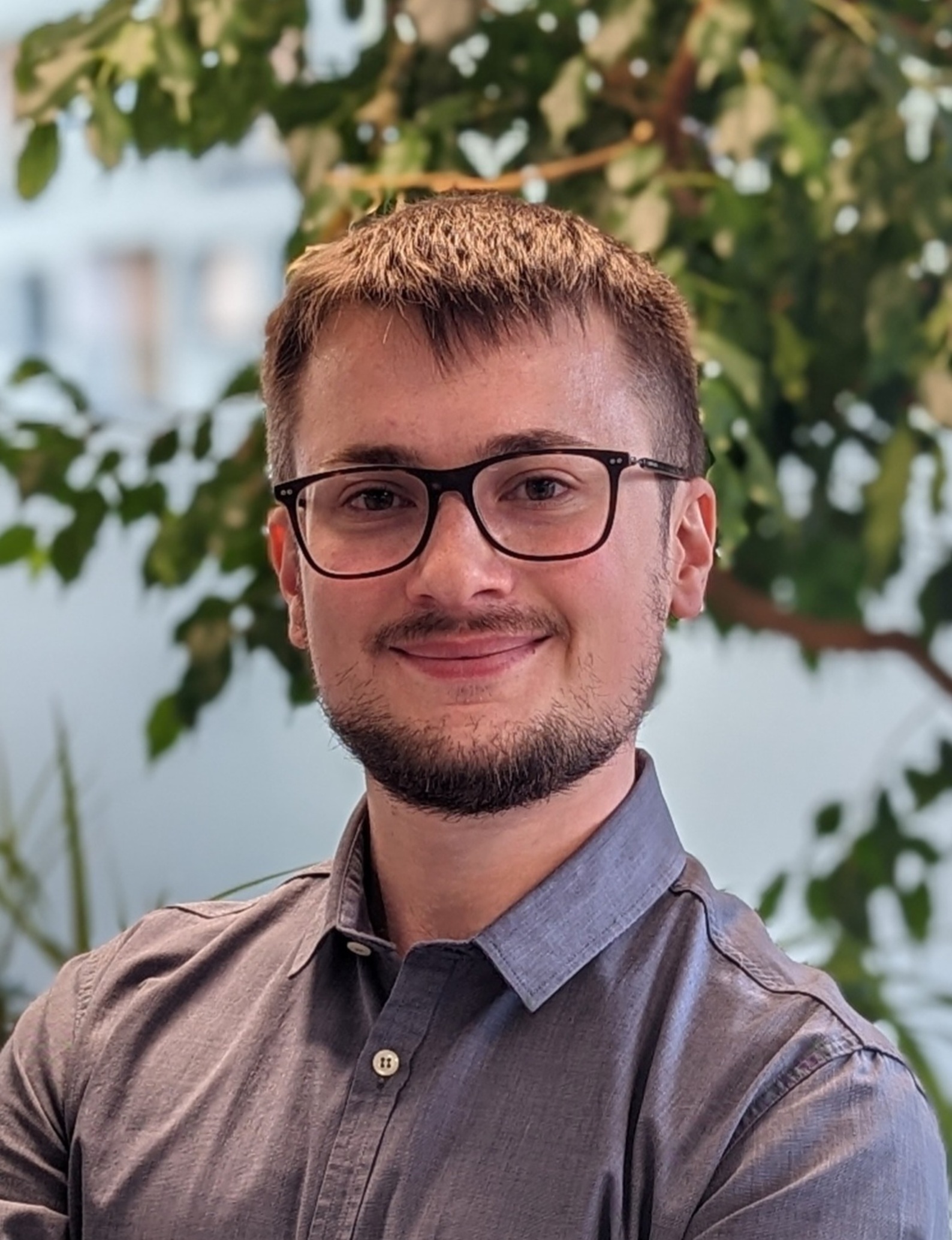}}]{Luca Mattarolo} fulfilled his Master's studies in Mathematics at the Department of Mathematics, Computer Science and Physics, University of Udine, in Italy, in 2019. He is currently working as a Database Specialist at Secure Switzerland AG, taking care of the technical aspects of the power grids' data integration. His main field of interest is the analysis and development of algorithms that generate optimal or close-to-optimal solutions for decisional problems.
\end{IEEEbiography}

\vskip -2\baselineskip plus -1fil

\begin{IEEEbiography}
[{\includegraphics[width=1.025in,height=1.25in,clip]{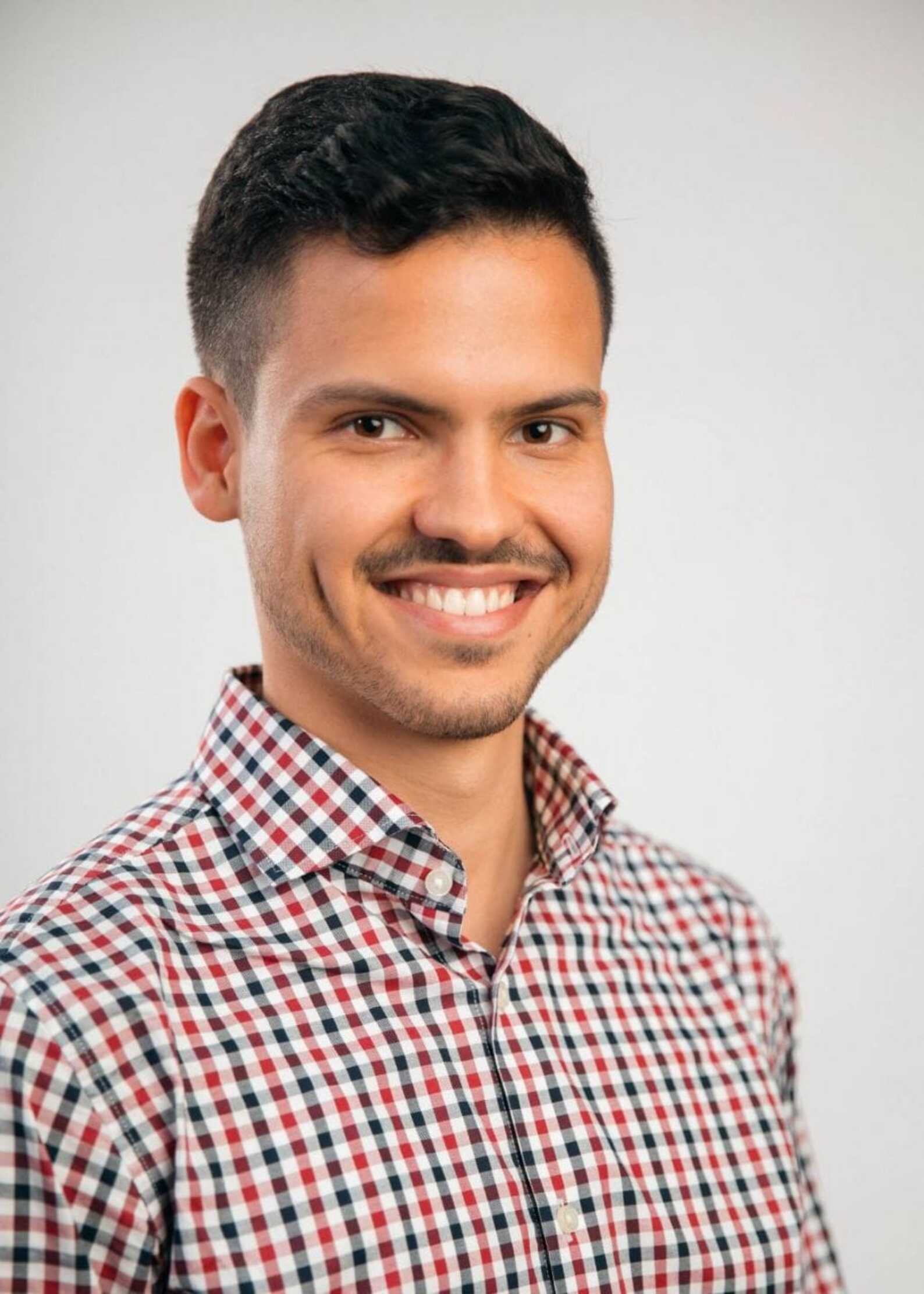}}]{Joel da Silva André} fulfilled his Bachelor's studies in Mechanical Engineering in 2017 and his Master studies in Energy Sciences at ETH Zürich at the Department of Information Technology and Electrical Engineering in 2021. In his Master's studies, his focus was on optimization in power systems. He is currently working as a Project Engineer at Secure Switzerland AG.
\end{IEEEbiography}

\vskip -2\baselineskip plus -1fil

\begin{IEEEbiography}
[{\includegraphics[width=1.05in,height=1.18in,clip]{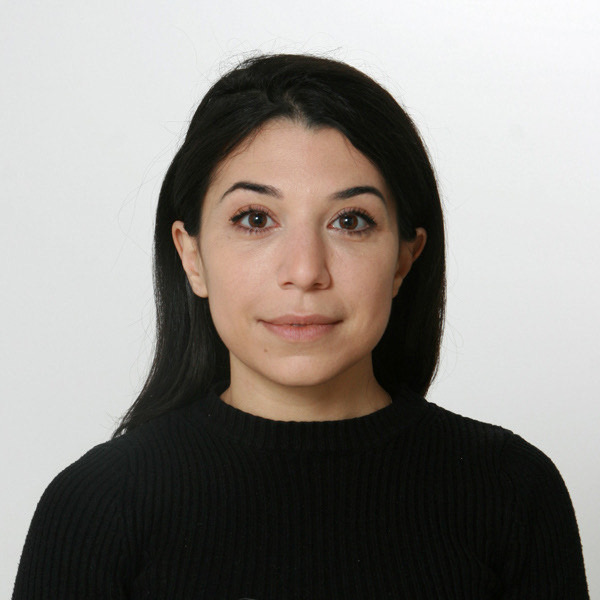}}]{Lydia Lavecchia} graduated with Honors from the University of Rome Tor Vergata with a Master of Science in Business Administration, with an emphasis in International Management and Marketing. Since 2017, Lydia has focused on developing her web development expertise. She worked in Secure Meters as a Full-Stack Software Engineer. Her main interest lies in full-stack web apps and high-impact growth strategies. 
\end{IEEEbiography}

\vskip -2\baselineskip plus -1fil

\begin{IEEEbiography}
[{\includegraphics[width=1in,height=1.25in,clip,keepaspectratio]{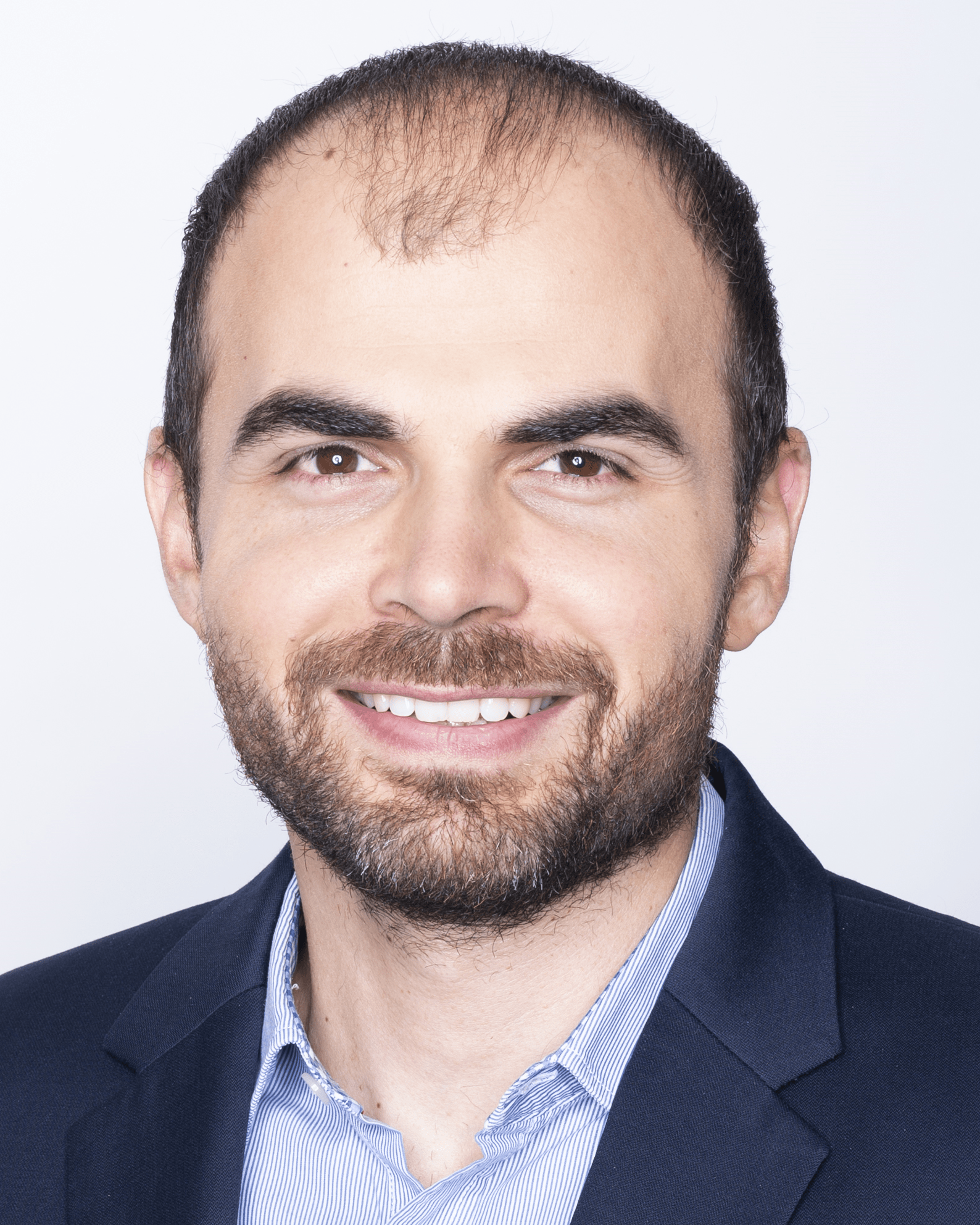}}]{Nikolaos Efkarpidis} fulfilled his Master's studies at the Department of Electrical and Computer Engineering, Aristotle University of Thessaloniki (AUTH), in Greece, in 2010 and received the Doctorate degree in Electrical Engineering from the Group of ESAT, KU Leuven, in Belgium, in 2016. He is currently working as Software Product Manager at Secure Switzerland AG. His main fields of interest include power quality control in power systems and the development of energy management algorithms at both network and end-user level.
\end{IEEEbiography}

\vskip -2\baselineskip plus -1fil

\begin{IEEEbiography}
[{\includegraphics[width=1in,height=1.25in,clip]{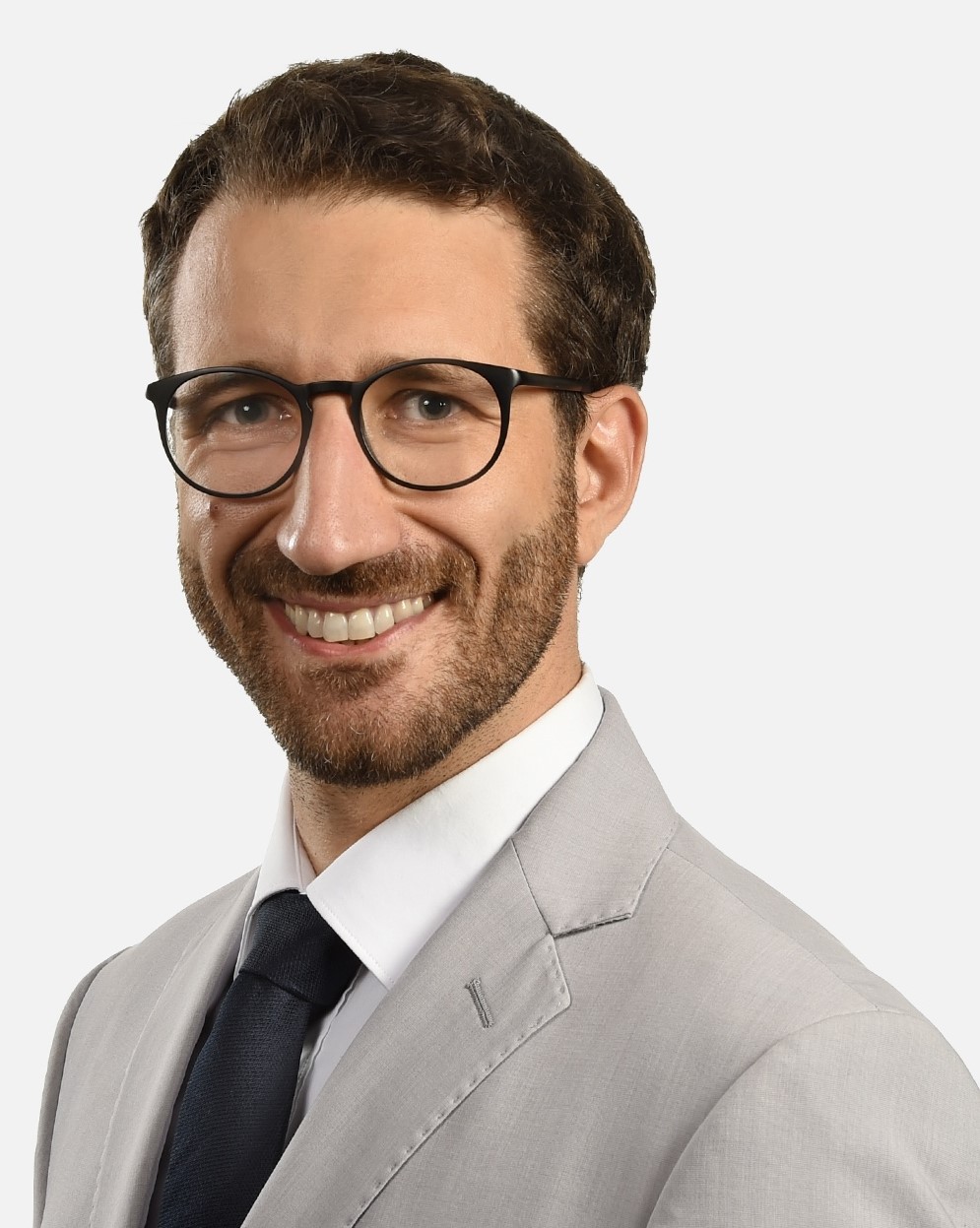}}]{Damiano Toffanin} is an electrical engineer who fulfilled his Master's studies at the Technical University of Denmark in 2016. He served as Head of Project Engineering at Secure Switzerland AG, developing simulation and monitoring software for distribution grids. His applied research focused on modeling residential loads and algorithms for grid diagnostics. Currently, he is pursuing a Master's in Business Administration at the University of St. Gallen.
\end{IEEEbiography}

\vfill

\end{document}